\RequirePackage{fix-cm}
\documentclass[smallextended]{svjour3}       
\smartqed  
\usepackage{graphicx}
\usepackage{lipsum}
\usepackage{pdfpages}

\usepackage{amsmath}
\usepackage{centernot}

\usepackage[flushleft]{threeparttable}

\usepackage{psfrag}
\usepackage{amsmath}
\usepackage{calc}
\usepackage{systeme}

\usepackage{mathtools}
\RequirePackage{textcomp}
\usepackage{mathrsfs}
\usepackage[utf8]{inputenc}
\usepackage[italian]{babel}
\usepackage[T1]{fontenc}
\usepackage{graphicx}
\usepackage{booktabs}
\usepackage{amsfonts}
\usepackage{amsmath}
\usepackage{textcomp}
\usepackage{subfigure}
\usepackage{mathtools}
\usepackage{mathrsfs}
\usepackage{mathtools}
\usepackage{cases}
\usepackage{cite}
\begin{document}

\title{A spectral approach to numerical simulations of the ADM equations
}
\subtitle{
}


\author{C. Meringolo$^1$  \and
      S. Servidio$^1$         \and
        P. Veltri$^1$ 
}


\institute{C. Meringolo \at
              \email{claudiomeringolo@hotmail.it} \\ 
              ~ \\
              $^1$  Dipartimento di Fisica, Universit\`a della Calabria, I-87036 Cosenza, Italy \\
              Tel.:+39 0984 49 6138\\
              Fax: +39 0984 49 4401\\
}

\date{Received: date / Accepted: date}

\maketitle

\begin{abstract}
We present a numerical study of the  Einstein equations, according to the Arnowitt-Deser-Misner (ADM) formalism, in order to simulate the dynamics of gravitational fields. We took in consideration the original $3+1$ decomposition of the ADM equations, in vacuum conditions, in simplified geometries. The numerical code is based on  spectral methods, making use of filtering (de-aliasing) techniques. The algorithm has been stabilized via an adaptive time-refinement, based on a procedure that checks self-consistently the regularity of the solutions. The accuracy of our numerical model has been validated through a series of standard tests. Finally, we present also a new kind of initial data that can be used for testing numerical codes.
\keywords{Numerical relativity \and ADM  \and numerical simulations}
\end{abstract}

\section{Introduction}
\label{intro}

In last years, many different numerical evolution schemes for Einstein
equations have been developed and proposed, in order to address stability and accuracy problems that
have interested the numerical relativity community for decades. Some of these
approaches have been tested on different spacetimes, and conclusions have
been drawn based on these tests. However, differences in results originate from
many sources, including not only formulations of the equations, but also gauges, boundary conditions, numerical methods and so on  \cite{bernuzz4, bernuz5,bernuz3,bernuz2,camp,naka,rezzol,prk, cap, ciufolini, baumg4, admyork, baumg3, 9, 13, 14, Bernuz}.
In this paper we present a $2+1$ formalism of the standard ADM decomposition \cite{ADM}, in vacuum condition, in order to solve Einstein equations. 
Our numerical code is based upon a spectral approach and periodic boundary conditions. The last ones are characteristic of homogeneous and localized regions of the space-time.
We use two different  anti-aliasing filters in order to minimize the aliasing instability due to nonlinear terms, and we show how a 'smoothed' filter leads to a more stable simulation than a truncated 'Heaviside' filter.
We further stabilize the code checking the right value of the time-step $dt$ of integration via an adaptive time-refinement, based on a procedure that checks self-consistently the regularity of the solutions, called $RSC~ condition$. 
We find a new solution of initial data that satisfies
the constraint equations, and leads to standing
waves of the metric tensor. These waves are sinusoidal for small amplitude of the perturbation, and become even more asymmetrical as the amplitude increases. 
The accuracy of our numerical model has been validated through a series of standard tests, like gauge wave tests and robust stability tests, as suggested by Alcubierre, Dumbser, Rezzolla \textit{et al.} \cite{Alc, Dum, bab, bab2, Brown}.
The code successfully passed various numerical tests, showing accuracy, stability and robustness.

\section{Formulation}
Throughout this paper, Latin indices are
spatial indices and run from 1 to 3, whereas Greek indices
are spacetime indices and run from 0 to 3.
The basic equation is the vacuum Einstein equation:
\begin{equation}
G_{\mu \nu}=0.
\label{einstein}
\end{equation}
To solve Eq.~(\ref{einstein}) we start from the standard $3+1$ formulation of general
relativity of Arnowitt, Deser, and Misner \cite{book, Alcu}, and write the line element as:
\begin{equation}
\nonumber
ds^2 = -\alpha^2 dt^2 +\gamma_{ij} \big(dx^i+\beta^i dt\big)\big(dx^j+\beta^j dt\big),
\end{equation}
where $\alpha$, $\beta^k$ and $\gamma_{ij}$ are the lapse function, the shift vector
and the spatial metric, respectively \cite{book,Alcu}. Using
the 3+1 formalism, the Einstein equation is split into the
constraint equations and the evolution equations. The
Hamiltonian and momentum constraints are:
\begin{subequations} 
	\label{constraints}
	\begin{equation}
	R+  K^2  -  K_{ij} K^{ij} = 0,
	\end{equation} 
	\begin{equation}
	D_i\big(K^{ij} -\gamma^{ij} K\big) = 0,
\end{equation}
\end{subequations}
where $K_{ij}$, $K$, $R$ and $D_i$ are the extrinsic curvature,
the trace part of $K_{ij}$, the scalar curvature of a 3D hypersurface and the covariant derivative with respect of
$\gamma_{ij}$, respectively. The set of equations (\ref{constraints}) represent the initial data problem, discussed in detail in Ref.\cite{baumg2, init, york1}. The evolution equations for the spatial metric and the
extrinsic curvature are, respectively, written as:

\begin{subequations} 
	\label{evot} 
	\begin{equation}
	(\partial_t - \mathcal{L}_\beta) \gamma_{ij} = -2 \alpha K_{ij},\\ \label{eq:prima3}
	\end{equation} \vspace{1pt}
	\begin{equation}
	\begin{split}
	(\partial_t - \mathcal{L}_\beta) K_{ij}= \alpha \big(R_{ij}-2K_{ik}K^{k}_{~j}+KK_{ij}\big)
	-D_i D_j \alpha.\\
    \label{eq:seconda}
	\end{split}
	\end{equation}
\end{subequations}
In what follows we will restrict to
the case of zero shift ($\beta^k =0$). The ADM evolution equations then reduce to:

\begin{subequations}  
	\begin{equation}
	\partial_t \gamma_{ij} = -2 \alpha K_{ij},
	\end{equation}
	\begin{equation}
	\partial_t K_{ij}= \alpha \big(R_{ij}-2K_{ik}K^{k}_{~j}+KK_{ij}\big)
	-D_i D_j \alpha.
	\end{equation} 
\end{subequations}
Finally, we need evolution equation for the lapse $\alpha$, i.e., we need to choose a slicing condition.
In the Bona-Massó (BM) formalism \cite{ bona1,bona2,bona3,BM} the following slicing condition is used:
\begin{equation}
\partial_t \alpha = - \alpha^2 f(\alpha) K,
\label{slicing}
\end{equation}
with $f(\alpha)>0$ but otherwise arbitrary\cite{ gau4, gau, gau3, baumg1}.

\subsection{A 2+1 ADM formulation}
In this paper, we will study a simplified, reduced
geometry. In particular, we now introduce a $2 + 1$ decomposition, in vacuum
conditions and zero shift.
For example, one can assume that at $t=0$ the 3-dimensional metric and the extrinsic curvature are of the type:
\begin{equation}
\label{2de}
\gamma_{ij}=
\begin{pmatrix}
\gamma_{xx} &\gamma_{xy}  & 0 \\
\gamma_{yx} & \gamma_{yy}  & 0 \\
0 & 0  & 1 \\
\end{pmatrix}
,~~~K_{ij}=
\begin{pmatrix}
K_{xx} & K_{xy}  & 0 \\
K_{yx} & K_{yy}  & 0 \\
0 &  0 & 0 \\
\end{pmatrix},
\end{equation}
with the extra-condition that nothing depends of $z$: 
\begin{equation}
\label{hhhq}
\partial_z \gamma_{ij}=\partial_z K_{ij}=0.
\end{equation}
It is clear that these initial conditions imply that all the $z-$components of the spatial Ricci tensor $R_{ij}$ are zero:
\begin{equation}
\label{gud}
R_{zj}=0,~~~\forall j.
\end{equation}
This hold at $t=0$ but conditions (\ref{2de}), (\ref{hhhq}) and (\ref{gud}) guarantee that the terms of the type $R_{zj}$ do not mix with the $R_{xx},R_{xy}$ and $R_{yy}$ terms during the time evolution. This happens for both the Ricci tensor and for all the other dynamic variables.

Then it is possible rewrite the evolution equations for $2D$ quantities formally in the same way of (\ref{evot}), except that indices run over two possible values, that is $(i, j)= x,y$.
This proves that a $2+1$ subcase is totally consistent with the full $3+1$ case.

\section{Numerical technique}

\subsection{Spectral method}

All of our numerical computations are carried out using
pseudospectral methods \cite{FFFT}. A brief outline of our method is as follows: given a system of partial differential equations 
\begin{equation}
\nonumber
\partial_t u(\bf x,\it t)= f[u(\bf x \it,t), \partial_i u(\bf x,\it t)],
\end{equation}
where $u$ is a collection of dynamical fields (i.e. $\gamma_{ij}, K_{ij},\alpha$), the solution
$u(\bf x ,\it t)$ is expressed as a time-dependent, truncated linear combination $u_N(\bf x,\it t)$
of spatial basis functions $\phi_k (\bf x \it)$, $k \in \mathbb{Z}$:
\begin{equation}
u(\bf x,\it t) \simeq u_N(\bf x,\it t)= \sum _{k=-n/2} ^{n/2}  \widetilde{u}_k(t) ~\phi_k (\bf x \it).
\label{fu}
\end{equation}
Associated with the basis functions is a set of $N$ collocation points $x_i$. Given spectral coefficients $\widetilde{u}_k(t)$, the function values at the collocation points $u_N(\bf x_{\it i},\it t)$ are computed
using Eq.~(\ref{fu}). Conversely, the spectral coefficients are
obtained by the inverse transform:
\begin{equation}
\nonumber
\widetilde{u}_k(t) = \sum _{i=-n/2} ^{n/2}  w_i~u_N(\bf x_{\it i},\it t) ~\phi_k (\bf x_{\it i} \it),
\end{equation}
where $w_i$ are weights specific to the choice of basis functions and collocation points. Thus it is straightforward to
transform between the spectral coefficients $\widetilde{u}_k(t)$ and the
function values at the collocation points $u_N(\bf x_{\it i})$.

Because the tests discussed here are periodic in all
spatial dimensions, we use \textit{Fourier basis} functions 	
$
\phi_k(\bf x \it) = e^{i\bf k \cdot x}
$.
Eq.~(\ref{fu}) then becomes:
\begin{equation}
\nonumber
u_N(\bf x,\it t) = \sum _{k=-n/2} ^{n/2}  \widetilde{u}_k(t) ~e^{i\bf k \cdot x}.
\end{equation}
Note that in a periodic domain, it is easy to show that the above series is simply truncated to $n=N/2$. Note also that the reality conditions gives $\widetilde{u}_k = \widetilde{u}^*_{-k}$.
To solve the differential equations, we evaluate spatial
derivatives analytically using the known derivatives of the
basis functions:
\begin{equation}
\nonumber
\partial_i u_N(\bf x,\it t) = \sum _{k=-n/2} ^{n/2}  \widetilde{u}_k(t) ~\partial_ie^{i\bf k \cdot x}.
\end{equation}
For smooth solutions, the spectral approximation
Eq.~(\ref{fu}) converges exponentially (error $\sim e^{-\lambda N}$ for some
$\lambda > 0$ which depends on the solution). This is much faster
than the polynomial convergence (error $\sim 1/N^p$) obtained
using $p$th-order finite-differencing  \cite{ggg}.

\subsubsection{The aliasing phenomenon and the anti-aliasing filters} \label{filtri}
Let consider now the simplest possible nonlinearity, i.e. the product of two functions $f(x)$ and $g(x)$ in a one-dimensional problem, defined by their truncated Fourier series with $N$ modes:
\begin{equation}
f_N(x) =  \sum _{p=-m} ^{m} \widetilde{f}_p~ e^{ip x},~~~~~~g_N(x) =  \sum _{q=-m} ^{m} \widetilde{g}_q~ e^{iq x},
\label{prod1}
\end{equation}
where $\widetilde{f}_p$ and $\widetilde{g}_q$ are the complex related Fourier coefficients.

The product of two functions in the physical space, defined of a finite, periodic grid, where therefore $m=N/2$, becomes a convolution product in Fourier space, namely
\begin{equation}
f(x)g(x) =  \Big(\sum _{p=-m} ^{m} \widetilde{f}_p~ e^{i p x} \Big)  \Big(\sum _{q=-m} ^{m} \widetilde{g}_q~ e^{iq x}\Big)= 
\sum _{p=-m} ^{m}\sum _{q=-m} ^{m} \widetilde{f}_p~\widetilde{g}_q ~ e^{i (p+q) x}.
\label{prod2}
\end{equation}
It can be clearly seen that the product (\ref{prod2}) contains high order harmonics respect to the truncated Fourier series of singles functions (\ref{prod1}), which cannot be represented on the initial grid. They will contribute to the well-know \textit{aliasing error} \cite{kstar},  since the Fourier transform of this product gives:
\begin{equation}
\widetilde{Q}_k = \int f(x)g(x) e^{-ikx }  d x =  \dots= \sum_{p+q=k} {\widetilde f}_p {\widetilde f}_q, 
\label{prod3}
\end{equation}
where we have defined a single sum over some selected couplings, namely $\sum_{p+q=k}\{...\} \equiv \sum_p \sum_q \{...\} \delta_{p+q,k}$. Now suppose that both $f$ and $g$ have all the harmonics, from $-N/2$ to $N/2$. One can immediately see how  products like Eq.~(\ref{prod3}) proliferates energy into $k>N/2$, causing the aliasing error and hence numerical instabilities. In order to eliminate instabilities due to aliasing in the quadratic nonlinear terms, it is then useful to define a $k^*$ in such a way that  all coefficients with $p,q,...>k^*$ are zero. For a quadratic nonlinearity of the type in Eq.s~(\ref{prod2})--(\ref{prod3}), it has been demonstrated that is sufficient to filter out modes with $k> k^*= 2N/3$, instead of $k^*=N/2$. This fully eliminates the aliasing instability (see e.g. \cite{canuto}). 

In summary, the technique is very simple: on the final product it is enough to set ${\widetilde Q}_k = 0$ for $k>2N/3$.
In the case of the Einstein field equations, it is important to consider the very high nonlinearity of the system. Let's take for example Eq.~(\ref{eq:seconda}), where one has products of the type:
\begin{equation}
\nonumber
\partial_t K_{xx} \sim ....+ \frac{1}{4}\alpha \gamma^{yx}(\partial_y \gamma_{xx})\gamma^{yy}(\partial_x\gamma_{yy}) + ...
\end{equation}
The above quantity has a nonlinearity of order 5, which corresponds to a terrible convolution in the Fourier space. Following the above decomposition, indeed, for a generic quintic product, in 2D, one has to consider generally 
\begin{equation}
\nonumber
Q({\bf x})= f_1({\bf x})f_2({\bf x})f_3({\bf x})f_4({\bf x})f_5({\bf x}).
\end{equation}
Taking the Fourier transforms, and applying the space-integral, one gets:
\begin{equation}
\nonumber
\widetilde{Q}_{\bf k} = \sum_{{\bf p}+{\bf q}+{\bf l}+{\bf h}+{\bf n}={\bf k}}   \widetilde{f}_1({\bf p})\widetilde{f}_2({\bf q})\widetilde{f}_3({\bf l})\widetilde{f}_4({\bf h})\widetilde{f}_5({\bf n}).
\end{equation}
It is easy to envision a process in which the above products produce immediately high-order harmonics and therefore a pronounced aliasing instability. 

Because of the above discussion, in this paper, different values of $k^*$ have been chosen, depending on the difficulty of the simulation and on the initial conditions type. Generally, by filtering high Fourier modes, the price to pay is the loss of effective resolution (information). However, by suppressing this high-$k$'s activity, the codes become more stable and accurate since the convolution in the quintic products does not push energy outside from the allowed $k$--space.

We will adopt Fast Fourier Transforms (FFT) \cite{FFT} to compute  spatial derivatives. For any product (and in general for any variable), we filter out the highest harmonics using two types of filter $\Phi_{k^*}(k)$. We will have then: 
\begin{equation}
\nonumber
f_N(x) = \sum_{k=-N/2}^{N/2} {\widetilde f}_k e^{i k x} \Phi_{k^*}(k).
\end{equation}
A first truncated 'Heaviside' filter is defined as:
\begin{subequations}
\begin{equation}
	\Phi_{k^*}(k)= 1 ~~if~|k|\leq k^*,
		\end{equation}
		\begin{equation}
	\Phi_{k^*}(k)= 0 ~~if~|k|> k^*,
	\end{equation}
	\label{filteralias2}
\end{subequations}
and a second smoothed filter is given by:

\begin{equation}
	\Phi_{k^*}(k)= e^{-a |\xi ^a|},
		\label{filteralias1}
\end{equation}
where $\xi = \frac{|k|}{k^*}$ and $a=30$.

In our numerical experiments we will use $k^* = \infty$ (no-filter), $k^*=N/2$ (grid-size), $k^*=N/3$ (typical quadratic nonlinearities),  $N/4$ and so on.

\subsection{Runge-Kutta method}
For our numerical tests we use a second-order Runge-Kutta (RK) method \cite{nrecipes}. Consider a function $y(t)$, with $t \in [0, +\infty)$ and $F(t,y)$ a generic function in the Cauchy problem:
\begin{equation} 
	\label{de}
	\partial_t\, y(t)=F(t, y(t)),~~~~
	y(t_0) = y_0.
\end{equation}
Assuming the function $y_n$ know at the discrete time interval $n$, the idea is to write the function $y_{n+1}$ at time $n+1$ as linear combination of the form:
\begin{equation}
y_{n+1}= y_n + a K_1(t_n, y_n)+b K_2(t_n, y_n)+O(\Delta t^3)
\label{runge1}
\end{equation}
where
\begin{equation}
\nonumber
K_1 = \Delta t \, F(t_n, y_n),~~~~~K_2=\Delta t \,F(t_n + \vartheta \Delta t, y_n + \varphi K_1),
\end{equation}
$\Delta t$ is the time-step of integration and $a, b, \vartheta, \varphi$ are parameters. Using a Taylor expansion, it is easy to get:
\begin{equation}
\nonumber
K_2 \simeq \Delta t \bigg[  F(t_n, y_n) + \vartheta \Delta t ~\partial_t F\Big|_{t_n, y_n}  +\varphi K_1~ \partial_y F\Big|_{t_n, y_n} \bigg] +O(\Delta t^3).
\end{equation}
In Eq.~(\ref{runge1}) one obtains:
\begin{equation}
\nonumber
\begin{split}
y_n + \Delta t~F(t_n, y_n) +\frac{1}{2}\Delta t^2 ~\partial_t F\Big|_{t_n} + \frac{1}{2}\Delta t^2 ~\partial y F\Big|_{y_n}= y_n + a \Delta t F(t_n, y_n)+ 
\\+b \Delta t F(t_n, y_n)+ b \vartheta \Delta t^2 ~\partial_t F\Big|_{t_n}+b\varphi K_1 \Delta t ~\partial_y F\Big|_{y_n}.  
\end{split}
\end{equation}
In order to obtain a second-order scheme, it must be $a+b=1$, $b \vartheta = 1/2$ and $b\varphi = 1/2$, while the parameter $\vartheta$ is free. Finally, with these substitutions, one has:
\begin{equation}
\nonumber
y_{n+1}=y_n + \bigg(1-\dfrac{1}{2\vartheta} \bigg) \Delta t\, F(t_n, y_n)+\frac{1}{2\vartheta}\Delta t\, F(t_n+\vartheta \Delta t, y_n+\Delta t \,F(t_n, y_n)).
\end{equation}
Choosing $\vartheta = 1/2$ one obtains the second-order Runge-Kutta scheme:
\begin{equation}
\nonumber
y_{n+1}=y_n+\Delta t \,F \left(t_n + \frac{\Delta t}{2}, y_n + F(t_n, y_n) \frac{\Delta t}{2}\right).  
\end{equation}

\subsection{Running Stability Check (RSC)} \label{rsc1}
In numerical methods, the Courant-Friedrichs-Lewy (CFL) condition is a necessary convergence condition for the solution of certain partial differential equations
problems \cite{CFL}, in particular for explicit time integration schemes.
Suppose to describe the motion of a wave traveling through a discrete spatial
grid with speed $v$. One has to choose a length interval $\Delta x$ of spatial grid and a time
step integration $\Delta t$, but these two quantities are not really independent: in order
to obtain “good results”, the CFL condition imposes that it must be $\Delta t = f (\Delta x)$.
Schematically, the condition says that if a generic wave is moving across a discrete
spatial grid, then the time step must be less than the time for the wave to travel to
adjacent grid points. This means that time step and “space step” are tight related,
and in particular the “space step” (i.e. the grid point separation) fix an upper limit
for the time step: if the second one is reduced, the first one must also decrease.
For a simple one-dimensional propagating fluctuation, the CFL condition is
given by $\Delta t < C \frac{\Delta x}{v}$,
where $C$ is called Courant number, and in general is chosen to be $1/2$, in order to
better satisfy the condition.
A general criterion for determination of the time-step integration dt is necessary
in order to maximize the efficiency of the code and avoid instabilities, especially in
the unknown case of the ADM equations. For this purpose, we will elaborate the
CFL idea, in a more general sense. For any ADM dynamic variable, let say $A_{ij}$ ,
there is an evolution equation
\begin{equation}
\frac{\partial A_{ij}}{\partial t}\simeq \frac{\Delta A_{ij}}{\Delta t},
\label{fff}
\end{equation}
from Eq.~(\ref{fff}) one can estimate the time step relative to the dynamical variables $A_{ij}$ as 
$\mathcal{T}(A_{ij}) \stackrel{d}{=} \frac{A_{ij}}{\partial_t A_{ij}} $, and one obtains respectively for the metric, the extrinsic curvature and the lapse
\begin{subequations}
	\begin{equation}
	\nonumber
	\mathcal{T}(\gamma_{ij}) \stackrel{d}{=}- \frac{\gamma_{ij}}{2 \alpha K_{ij}},
	\end{equation}
	\begin{equation}
	\nonumber
	\mathcal{T}(K_{ij}) \stackrel{d}{=} \frac{K_{ij}}{\alpha \big(R_{ij}-2K_{ik}K^{k}_{~j}+KK_{ij}\big)
		-D_i D_j \alpha},
	\end{equation}
	\begin{equation}
	\nonumber
	\mathcal{T}(\alpha) \stackrel{d}{=} -\frac{\alpha}{ \alpha^2 f(\alpha) K}.
	\end{equation}
\end{subequations}
It is evident that variables that have a large time-derivative (they are fluctuating fast), have a small related $\mathcal{T}$ (less stable). Analogously, small time derivatives (or very large functions) lead to high $\mathcal{T}$ (more stable region). Furthermore, from our preliminary tests, it turns out that a small $\mathcal{T}$ anticipate the typical code-crashing.  The technique hence consist of an interesting  monitoring during the evolution of the code, together with the violation of the AMD constraints. Note also that the above technique it might provide a general guess for the choice of the integration time step in the second-order Runge-Kutta technique.
From the above reasoning, since there are several control times $\mathcal{T}$, one can choose the time step constrained to the following general  expression:
\begin{equation}
\Delta t < C~min \{ \mathcal{T}_j \}
\label{RSC}
\end{equation}
where $C$ is the Courant number and index $j$ run over all the ADM dynamic variables. This  method is called \textit{Running Stability Check} (RSC), since it computes continuously in time the minimum time step, controlling all the possible derivatives and variables. Eq.~(\ref{RSC}) allows to guess a good $dt$ during the numerical simulation, even in a self-adjusting fashion. 
Essentially, whenever the characteristic times $\mathcal{T}$ becomes too small, the code reduces its time step of the second order Runge-Kutta. This method is called \textit{adaptive time refinement} and leads to an improvement of stability, as will be shown in the next testbeds.

\section{Standard numerical testbeds}

In this Section we will perform new direct numerical simulations of the gravitational dynamics. We will explore all the standard numerical testbeds, suggested by Alcubierre, Dumbser, Rezzolla et al. \cite{Alc, Dum, bab, bab2, baumg2}, in order to validate our code. For each test, we will check accuracy by inspecting the conserved quantities of the ADM formalism, applying the running stability check and varying the anti-aliasing filters. An adaptive Runge-Kutta method will be employed in order to further ensure numerical stability. After these fundamental tests, we show a new possible initial condition that leads to standing nonlinear waves. 

\subsection{Robust stability test}

The robust stability testbed efficiently reveals exponentially growing modes which
otherwise might be masked beneath a strong initial signal for a considerable evolution time.
It is based upon small random perturbations of Minkowski space. 
 As suggested by Alcubierre \textit{et al.}~in Ref.~\cite{Alc}, the starting configuration is a flat Minkowski metric 
\begin{equation}
\nonumber
ds^2 = -dt^2 + dx^2 + dy^2.
\end{equation}
The procedure consists of adding to the above metric some random perturbations, distributed over every variable ($\gamma_{ij}, K_{ij}, \alpha$).
The idea is that if a code cannot stably evolve a random noise then it will be unable to
evolve a real initial data. In this test, the initial metric has been initialized as
\begin{equation}
\gamma_{ij}=\eta_{ij} +\varepsilon_{ij},
\label{noisee}
\end{equation}
where $\varepsilon_{ij}$ is the random perturbation (small random numbers generated via classical algorithms \cite{nrecipes}). 
The amplitude of the perturbation is small enough so that the evolution remains in the linear regime,  unless instabilities arise. This corresponds to the following choice:
\begin{equation}
\nonumber
\varepsilon \in \left[-{10^{-10}},+{10^{-10}}\right].
\end{equation}	
In all robust stability test the harmonic gauge was used [i.e., $f(\alpha)=1$ in the slicing equation (\ref{slicing})].
\\
We chose a spatial domain $x, y \in [ 0, 1 ]$, a spatial grid $N_x \times N_y = 64^2$, $dx=dy=2^{-6}$ and a $dt = 2 \cdot 10^{-3}$.
Since the initial
data violate the constraints, any instability can be expected to lead to an exponential growth of
constraints.	
In order to stabilize the code, the anti-aliasing filter, described by Eq.s~(\ref{filteralias2}) and (\ref{filteralias1}), can be used here to show whether it can improve the stability of the code.
\begin{figure}
\centering
\includegraphics[width=58mm]{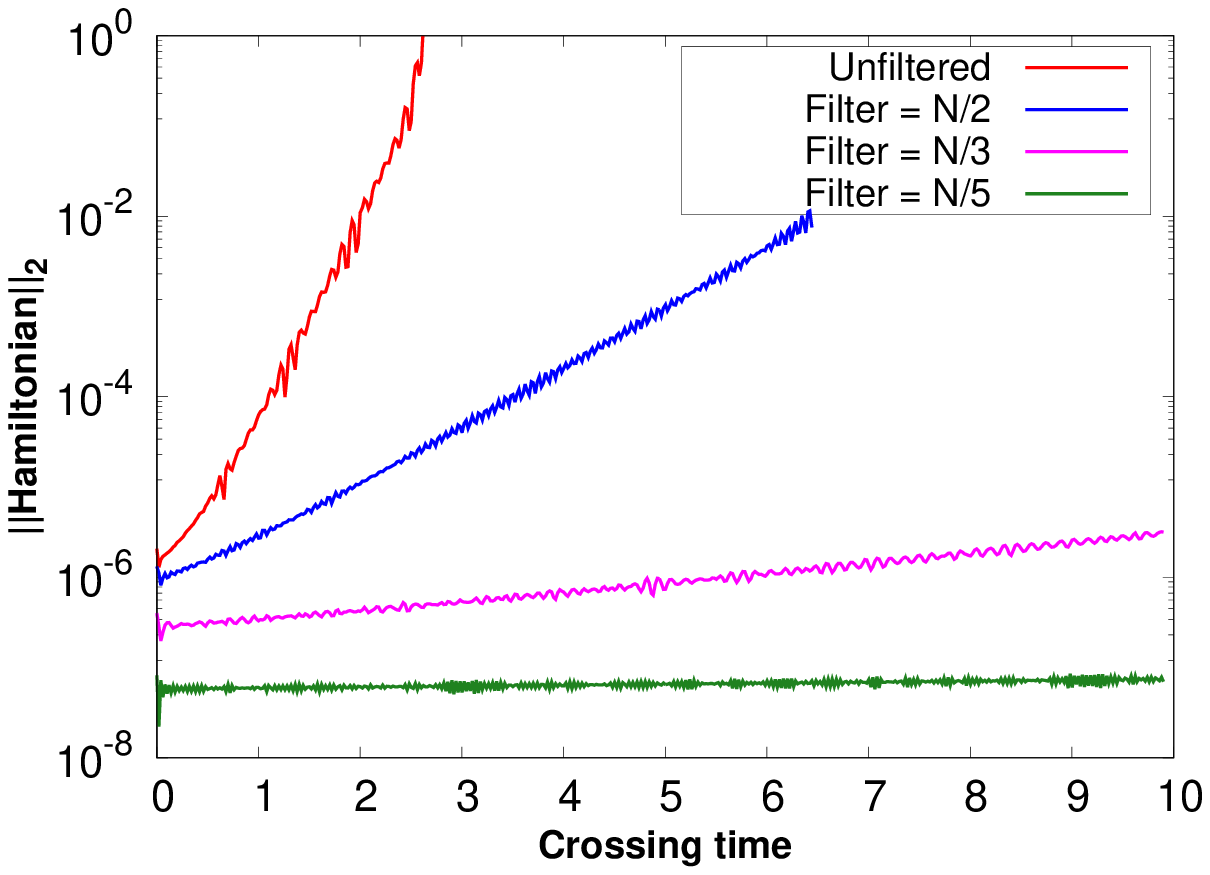}
\includegraphics[width=59mm]{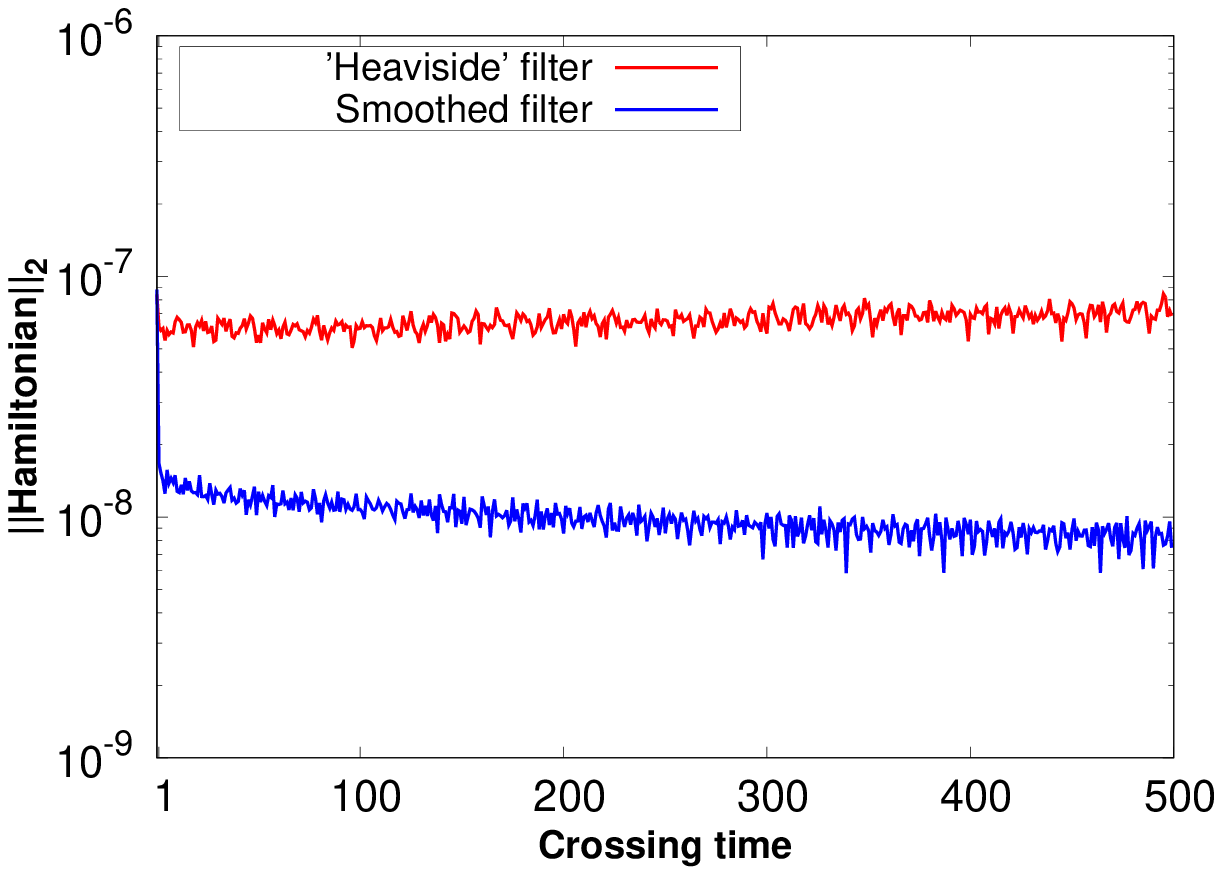}
\caption[ ]{\footnotesize{}Left: Hamiltonian constraint vs time for Minkowski flat space with random noise as perturbation, respectively unfiltered, with $k^*=N/2$, $k^*=N/3$ and $k^*=N/5$ 'Heaviside' anti-aliasing filter. These runs have been summarized in Table~\ref{table} as RUN$_1$, RUN$_2$, RUN$_3$ and RUN$_4$. Right: $L_2$ errors of Hamiltonian constraint for the same test using a $k^*=N/5$ for both anti-aliasing filters described by Eq.s~(\ref{filteralias2}) and (\ref{filteralias1}), and $dt=5\cdot 10^{-4}$. Note that the smoothed filter better stabilize the code. The evolution is carried out for $t=500$, and these runs are reported in Table~\ref{table}, as RUN$_5$ and RUN$_6$.  }
\label{fig:31}
\end{figure}

The test runs for a time of $t = 10$ (corresponding to 10 crossing times)
or until the code crashes, and the performance has been monitored by looking at the evolution of the ADM constraints in time, i.e. by outputting the $L_2$ norm of the Hamiltonian constraint and the momentum constraints once per crossing. We define this measure as:
\begin{equation}
\Vert{L}\Vert_2=\sqrt{\dfrac{\int_{\Omega}\varepsilon^2\, d\Omega\,\sqrt{|\gamma|}}{\int_{\Omega}d\Omega\,\sqrt{|\gamma|}}},
\label{norml2}
\end{equation}
\\
where $\varepsilon$ denotes the local error of each of the ADM
quantities, i.e., Hamiltonian $H$ and momentum constraints
$M_i$, while $d\Omega\,\sqrt{|\gamma|}$ is the volume element.

As discussed in Section \ref{filtri}, since the ADM equations are strongly nonlinear, the $k^*=2N/3$ filter is not enough and then different filter was tested. The left panel of Fig.~\ref{fig:31} reports the evolution of the  $L_2$ norm of the Hamiltonian constraint, for the different filters (respectively unfiltered, with $k^*=N/2$, $ k^*=N/3$ and $ k^*=N/5$ filter) using the 'Heaviside' filter described by Eq.~(\ref{filteralias2}). These runs have been summarized in Table~\ref{table} as RUN$_1$, RUN$_2$, RUN$_3$ and RUN$_4$. It is obvious that the anti-aliasing filter improves noticeably the stability and the accuracy of the code.

In order to show the improvement of the smoothed filter and the robustness of the code, a last test with a $k^*=N/5$ using both filters described by Eq.s~(\ref{filteralias2}) and (\ref{filteralias1}) and a time step of $dt=5\cdot 10^{-4}$ has been performed, for $500$ crossing times. The $L_2$ norms of the Hamiltonian constraint are shown in the right panel of Fig~\ref{fig:31}. The test shows that the Hamiltonian remain essentially constant after a long time of simulation using the 'Heaviside' filter, while slightly decrease using the smoothed filter, emphasizing the goodness of our approach. These runs are reported in Table~\ref{table}, as RUN$_5$ and RUN$_6$.

\subsection{Gauge wave test}

The gauge-wave is a classical numerical recipe that tests how a code handles gauge dynamics.
As suggested again by Ref.~\cite{Alc}, the metric is given by:
\begin{equation}
ds^2 = -H(x,t)\, dt^2 +H(x,t) \,dx^2 +dy^2, 
\label{wavetest1}
\end{equation}
where $H(x,t) \stackrel{d}= 1-A~sin\big[2 \pi (x-t)\big]$ describes a sinusoidal gauge wave of amplitude $A$, propagating along the $x$-axis. Since derivatives are zero in the $y$
direction, the problem is essentially one-dimensional.
The metric (\ref{wavetest1}) implies $\beta^i=0$, and $K_{xx}=-\dfrac{\partial_t \gamma_{xx} }{2 \alpha}$.	
For the metric $\gamma_{ij}$ and the extrinsic curvature $K_{ij}$ one obtains respectively:
\begin{equation}
\nonumber
\gamma_{ij}=
\begin{pmatrix}
1-A~sin[2 \pi (x-t)] & 0  \\
0 & 1  \\
\end{pmatrix}
~,~~~~	K_{ij}=
\begin{pmatrix}
-A\pi\dfrac{ cos[2 \pi (x-t)]}{\sqrt{1-A~sin[2 \pi (x-t)]}} & 0  \\
0 & 0  \\
\end{pmatrix}
\end{equation}	
and one can easily demonstrates that these satisfy the initial data constraints in Eq.~(\ref{constraints}).	
	
This gauge-wave test was performed twice: once with a small amplitude $A$ in order to take the system in a linear regime
and a second test with a very large
amplitude $A$, as suggested by Dumbser \textit{et al.} \cite{Dum}.		
For the first run, a small amplitude $A=10^{-2}$ and a $k^*=N/3$ 'Heaviside' filter is used. 
We chose a spatial domain $x, y \in [ 0, 1 ]$, a spatial grid $N_x \times N_y = 64^2$, $dx=dy=2^{-6}$, a $dt = 5 \cdot 10^{-3}$ and the harmonic slicing ($f(\alpha)=1$).
However, in order to check the RSC conditions (see Section \ref{rsc1}) the test was performed twice: first one with a stationary time step, and a second one with an adaptive time refinement. These runs are summarized in Table~\ref{table} as RUN$_7$ and RUN$_8$.

\begin{figure}
\centering
\includegraphics[width=0.49\textwidth]{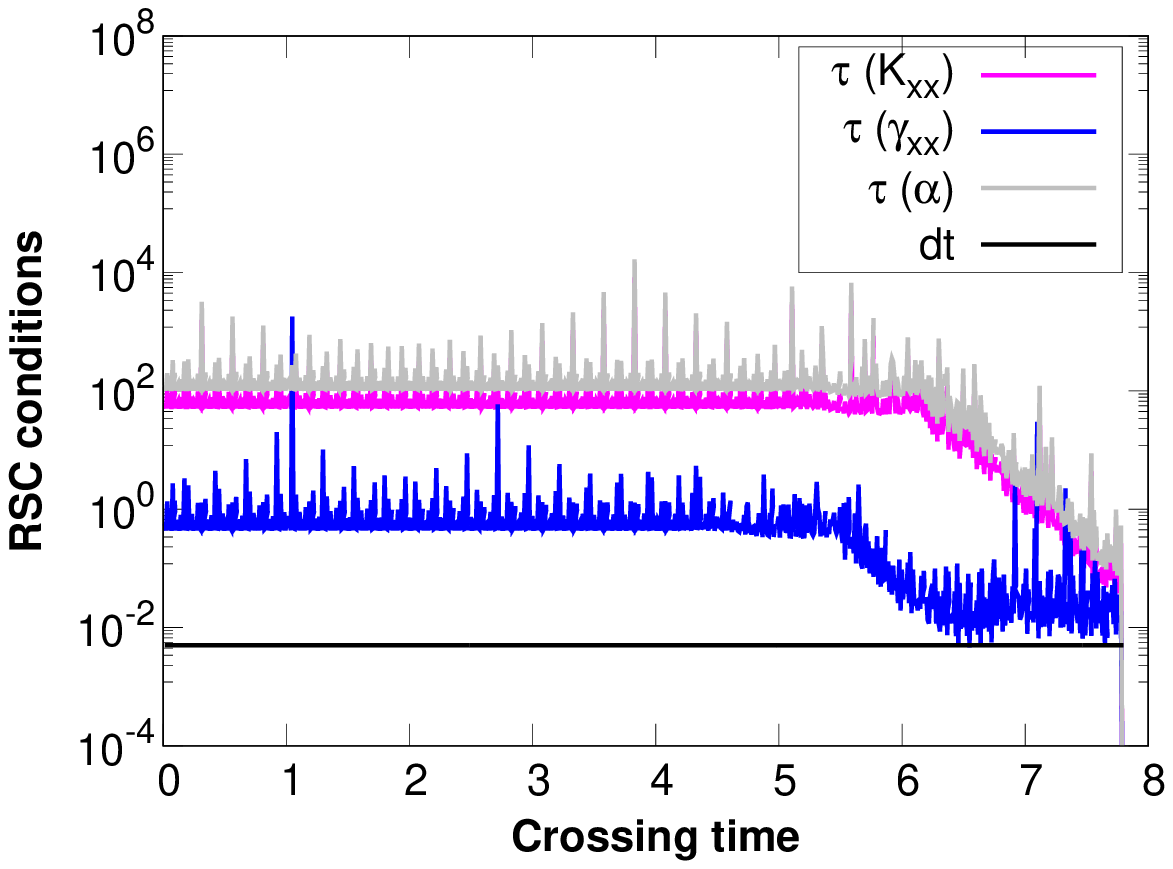}
\includegraphics[width=0.49\textwidth]{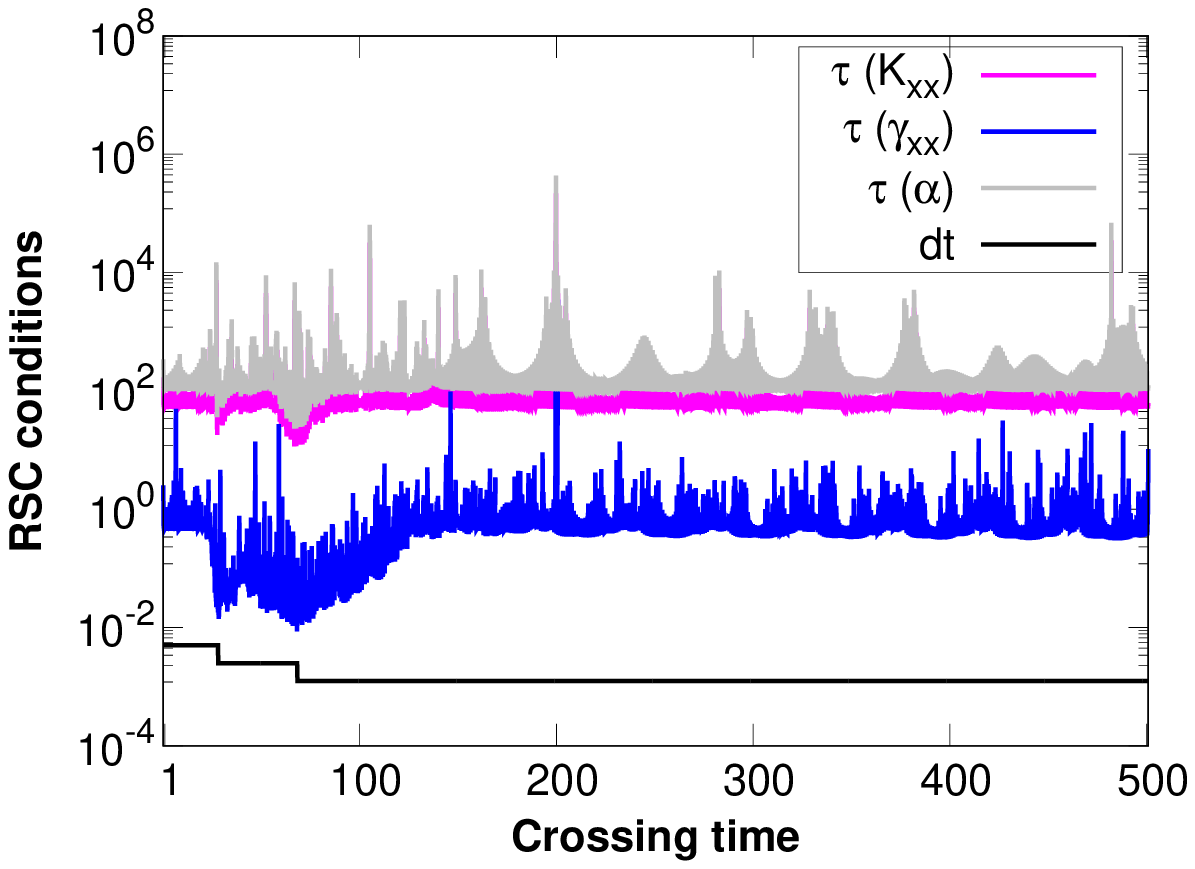}
\caption[RSC conditions and adaptive time refinement]{\footnotesize{}Small amplitude gauge wave test. Left: RSC conditions without adaptive time refinement: the code crashes at $t\simeq 8$. Right: RSC conditions with adaptive time refinement and $C = 1/4$: the code remain stable until $t=500$. For the symmetry of the problem, only $\mathcal{T}(K_{xx}), \mathcal{T}(\gamma_{xx}), \mathcal{T}(\alpha)$ have been reported. The full black line represents the time step of the simulation. These runs are reported in Table~\ref{table} as RUN$_7$ and RUN$_8$. }
\label{a1}
\end{figure}

Fig.~\ref{a1} (left) shows the time evolution of the RSC conditions in time. Note that for the symmetry of the problem, only $\mathcal{T}(K_{xx}), \mathcal{T}(\gamma_{xx}), \mathcal{T}(\alpha)$ have been reported. The full black line represents the time step of the simulation.
One can see that when one RSC condition becomes smaller than the time step, instabilities arises and the code crashes.
In the second test, it has been used an adaptive time refinement with $C=1/4$, and the run is carried out until 500 crossing times. The code is much more stable thanks to the fact that the time step remain ever below the RSC conditions, as shown in Fig.~\ref{a1} (right).

\begin{figure}  [t]
\centering
\includegraphics[width=0.50\textwidth]{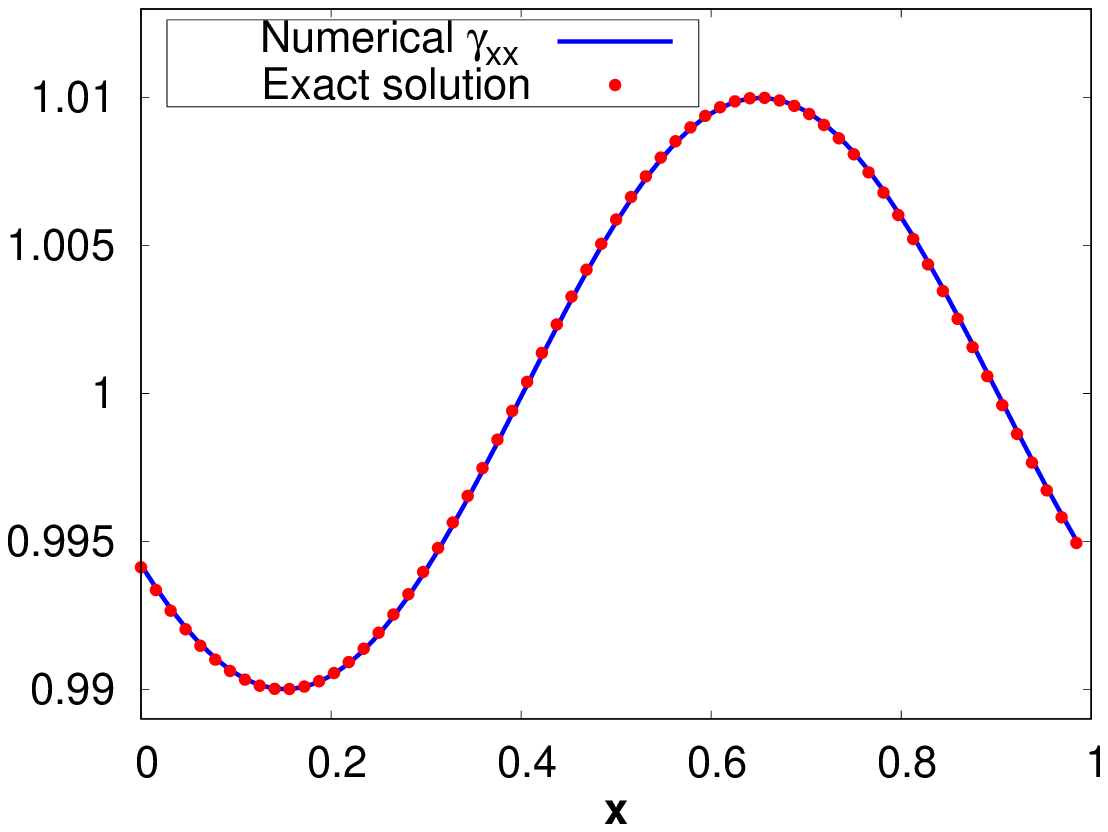}
\includegraphics[width=0.49\textwidth]{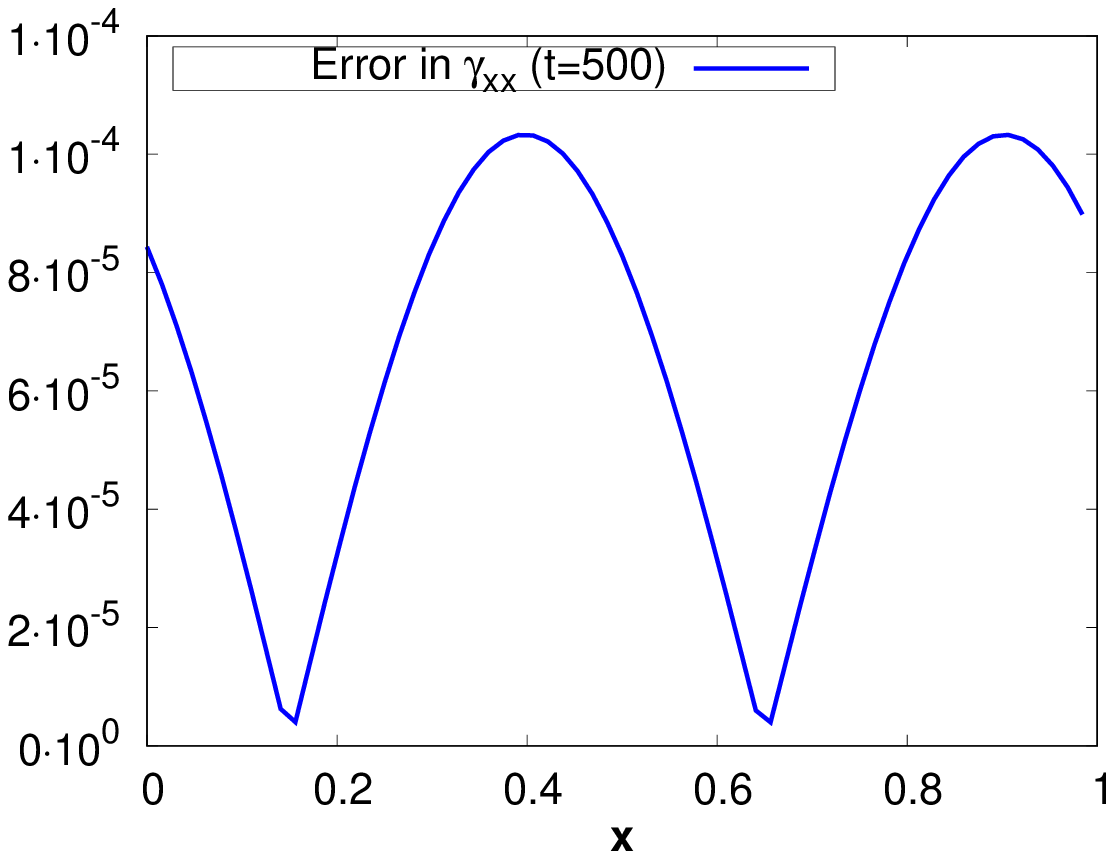}
\caption[Gauge wave test with small amplitude]{\footnotesize{}Left: Comparison of the wave form of $\gamma_{xx}$ with the exact solution, for a small amplitude perturbation $A^{-2}$, at $t=500$. Right: Difference between numerical $\gamma_{xx}$ and exact result at $t=500$. This test is reported in Table~\ref{table} as RUN$_9$.}	
\label{ap}
\end{figure}

Even if a code remains stable for long time, the test is meaningless without a good comparison between the numerical experiment an the analytic solution (if there is any). 		
In Fig.~\ref{ap} (left) we compare the wave-form of $\gamma_{xx}$, at $t=500$, with the exact solution. Here we use $dt=10^{-3}$. In Fig.~\ref{ap} (right) we report  the numerical error, in order to show the good consistency of the code, even if a $k^*=N/3$ 'Heaviside' filter is used. One can see that the error is two order of magnitude less than the amplitude of the wave. This test is reported in Table~\ref{table} as RUN$_9$.

Regarding the high perturbation amplitude, we have chosen at first $A=0.9$, with a square spatial grid of $n=128$ points and $dt=10^{-3}$. The other parameters are the same as used in the small amplitude case, including the slicing gauge. Again we have performed two experiments: first without anti-aliasing filter and second with $k^* = N/3$ smoothed filter defined in Eq.~(\ref{filteralias1}). 
A comparison between the nonlinear waveforms of the trace of the extrinsic curvature $K$ in both cases has been plotted in Fig.~\ref{ap}. As it can be seem there is a good improvement with the anti-aliasing filter. One can observe, indeed, an excellent agreement between the exact and the numerical solutions. In the unfiltered case, instead, numerical instabilities arise, as reported in the left panel of figure \ref{ap}. These runs are reported in Table~\ref{table} as RUN$_{10}$ and RUN$_{11}$, respectively.

\begin{figure} 
\centering
\includegraphics[width=0.49\textwidth]{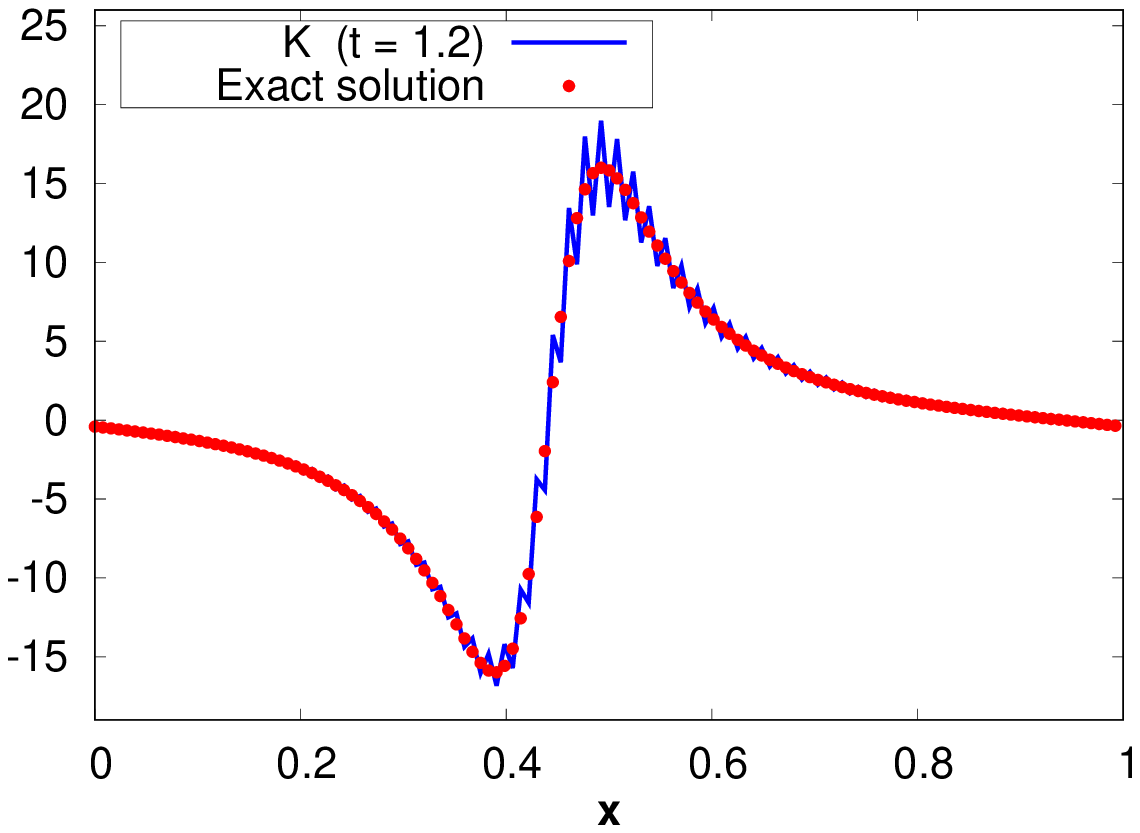}
\includegraphics[width=0.49\textwidth]{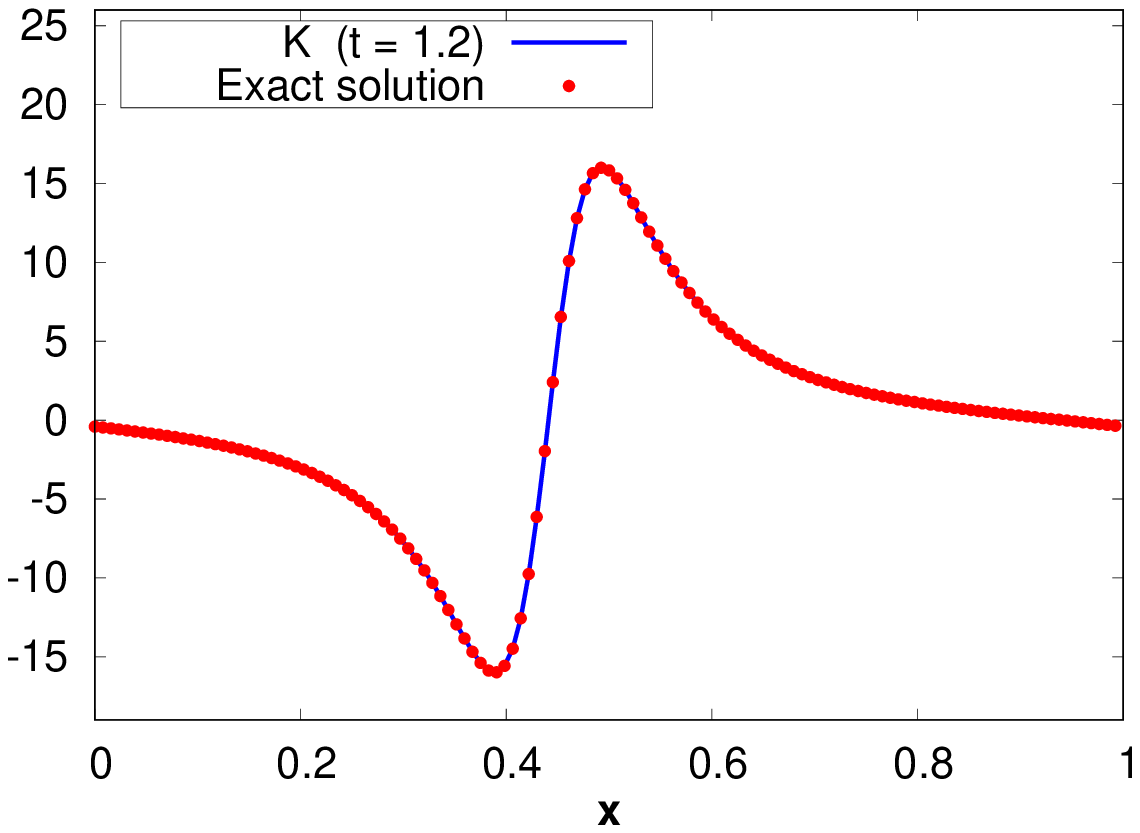}
\caption[]{\footnotesize{}Gauge-wave test case with amplitude A=0.9. Comparison of the wave form of trace of the extrinsic curvature K in nonlinear regime without filter (left) and  with $k^* = N/3$ smoothed filter (right) with the exact solution at $t=1.2$. The anti-aliasing filter leads to an improved wave form of $K$. These runs are summarized in Table~\ref{table} as RUN$_{10}$ and RUN$_{11}$.}
\label{p}
\end{figure}

A second test in nonlinear regime has been performed, this time with a larger perturbation $A=0.96$. Now a good choice for the filter has found to be $k^* = N/2.5$.
In order to test the different anti-aliasing filters we performed the test twice: the first one using the 'Heaviside' filter defined in Eq.~(\ref{filteralias2}), and the second one using the smoothed filter defined in Eq.~(\ref{filteralias1}).

\begin{figure}
\centering
\includegraphics[width=0.49\textwidth]{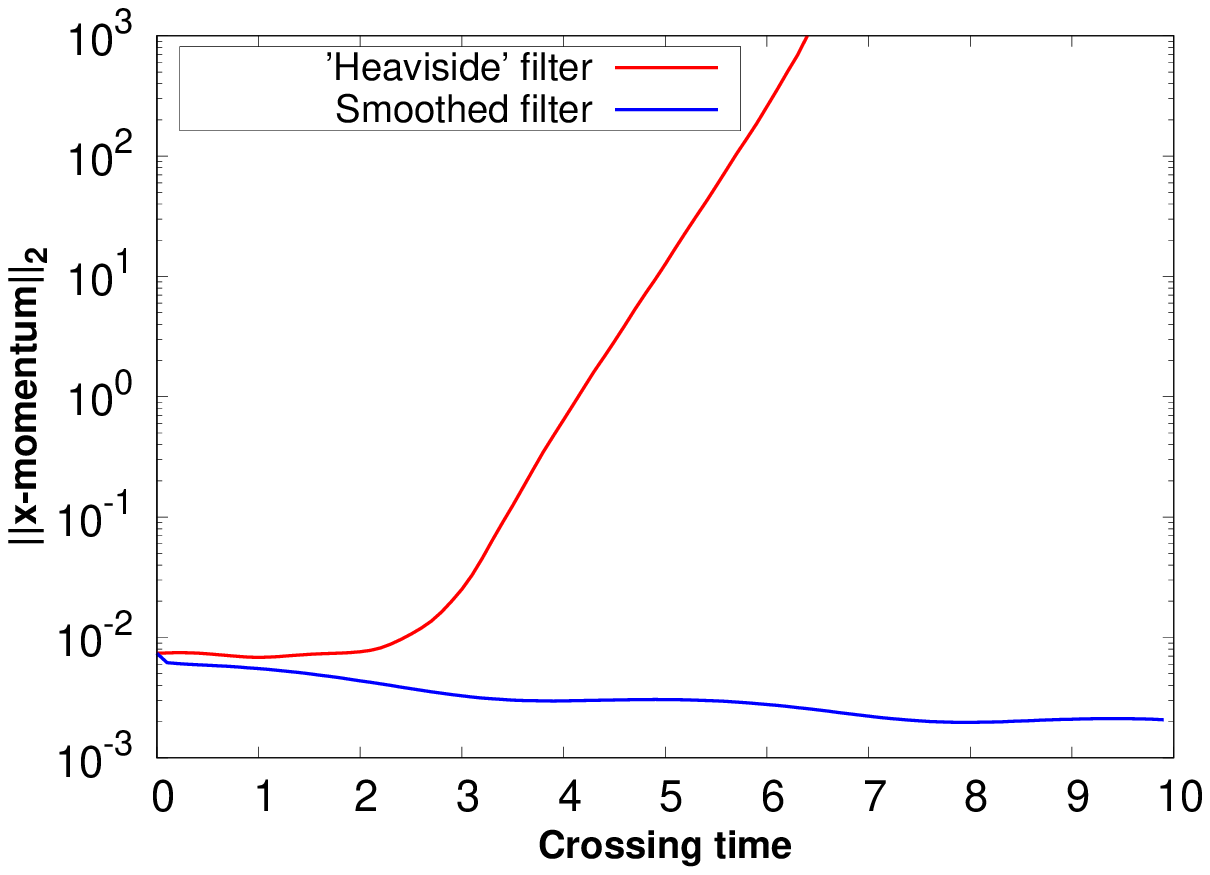}
\includegraphics[width=0.49\textwidth]{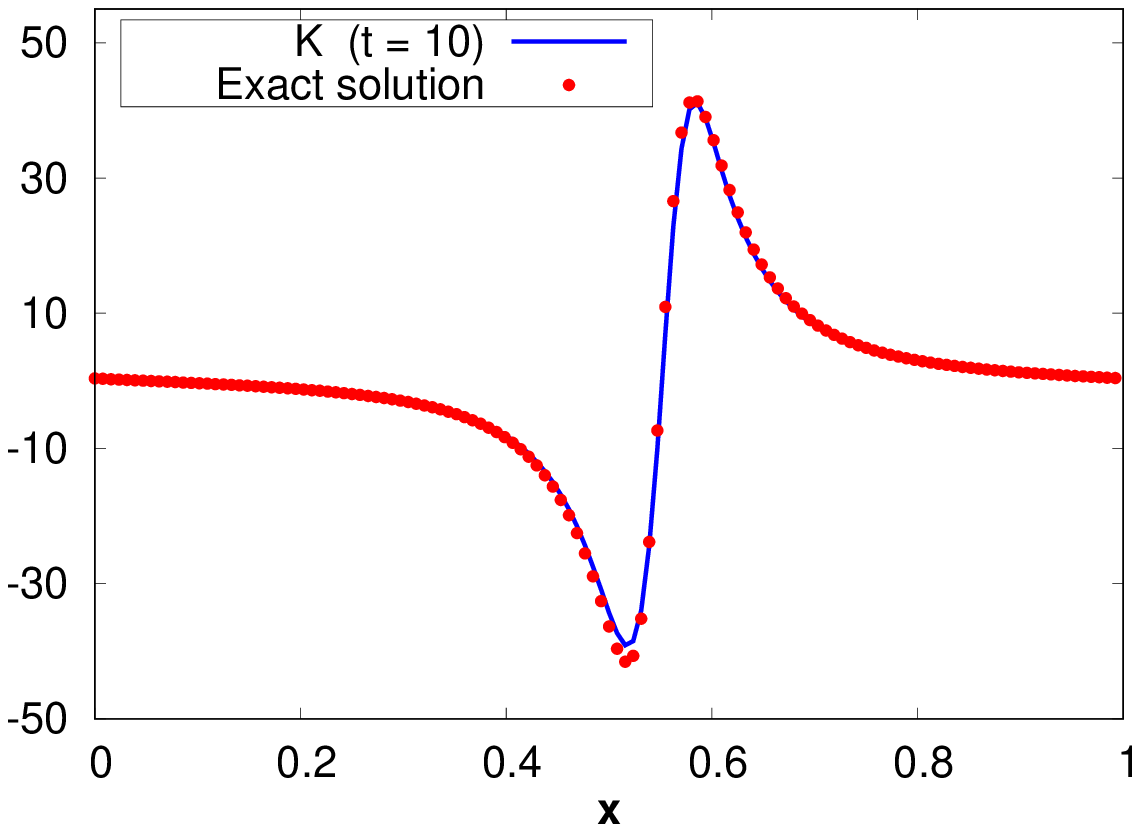}
\caption[]{\footnotesize{}Left: Evolution of $x-$momentum constraint for the gauge wave test with high amplitude $A=0.96$ using different filters described by Eq.s~(\ref{filteralias2},\ref{filteralias1}); with the 'Heaviside' filter the $x-$momentum constraint grows up and the code crashes, while with the smoothed filter the ADM constraint slightly decrease and the code stabilizes. The evolution has been carried out for $t = 10$ crossing times. These runs are reported in Table~\ref{table} as RUN$_{12}$ and RUN$_{13}$. Right: Comparison of the wave form of the trace of the extrinsic curvature $K$ with exact result and $A=0.96$ at t=10 using the smoothed filter. Note the good agreement despite the very high value of $A$.}
\label{hi}
\end{figure}

The time evolution of the $x-$momentum constraint for both tests is reported in Fig.~\ref{hi} (left), showing that the smoothed filter stabilize the code, while the 'Heaviside' filter leads to a growing of the ADM constraints and the code crashes. These runs are reported in Table~\ref{table} as RUN$_{12}$ and RUN$_{13}$.
In Fig.~\ref{hi} (right) we report the waveform of $K$ at time $t=10$, which is in excellent agreement with the exact solution.
It is important to emphasize that, with this amplitude, even if the system is in a very nonlinear regime, is still stable, thanks to the smoothed anti-aliasing filter.

\subsection{A new possible initial condition: the standing waves}

\begin{figure}[t]
\centering
\includegraphics[width=0.49\textwidth]{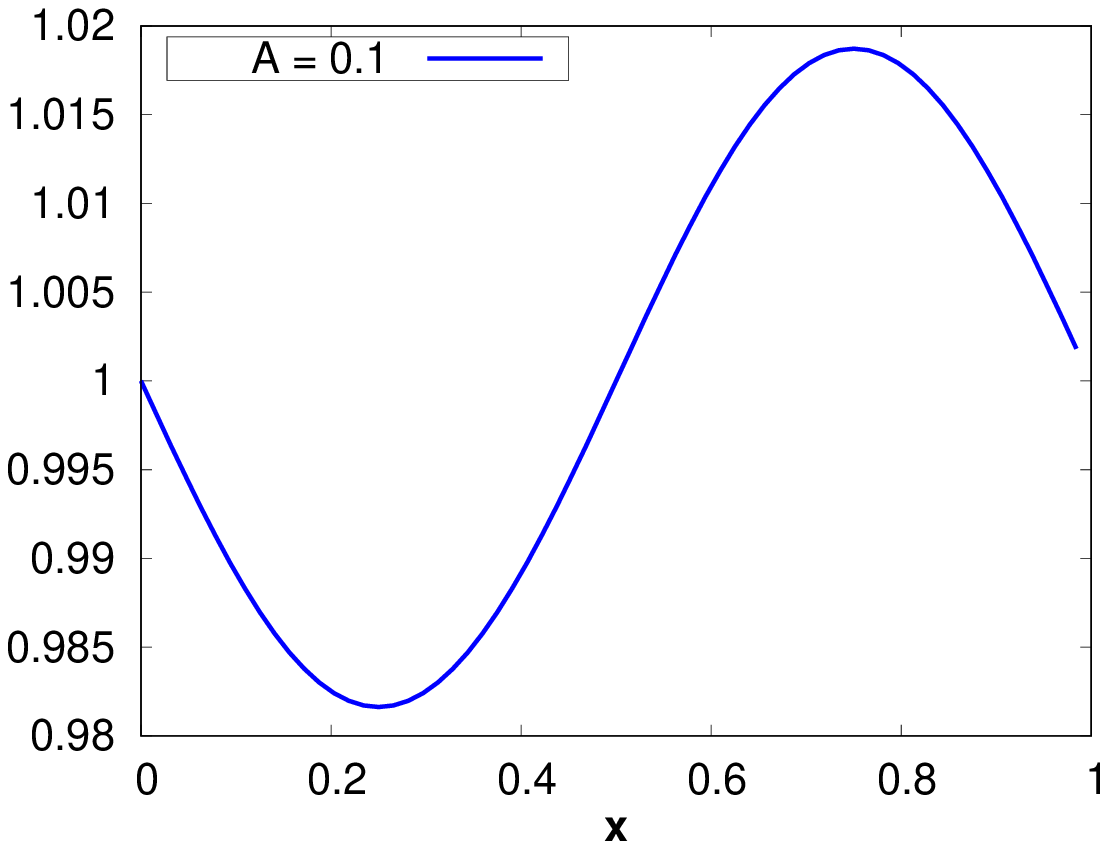}
\includegraphics[width=0.49\textwidth]{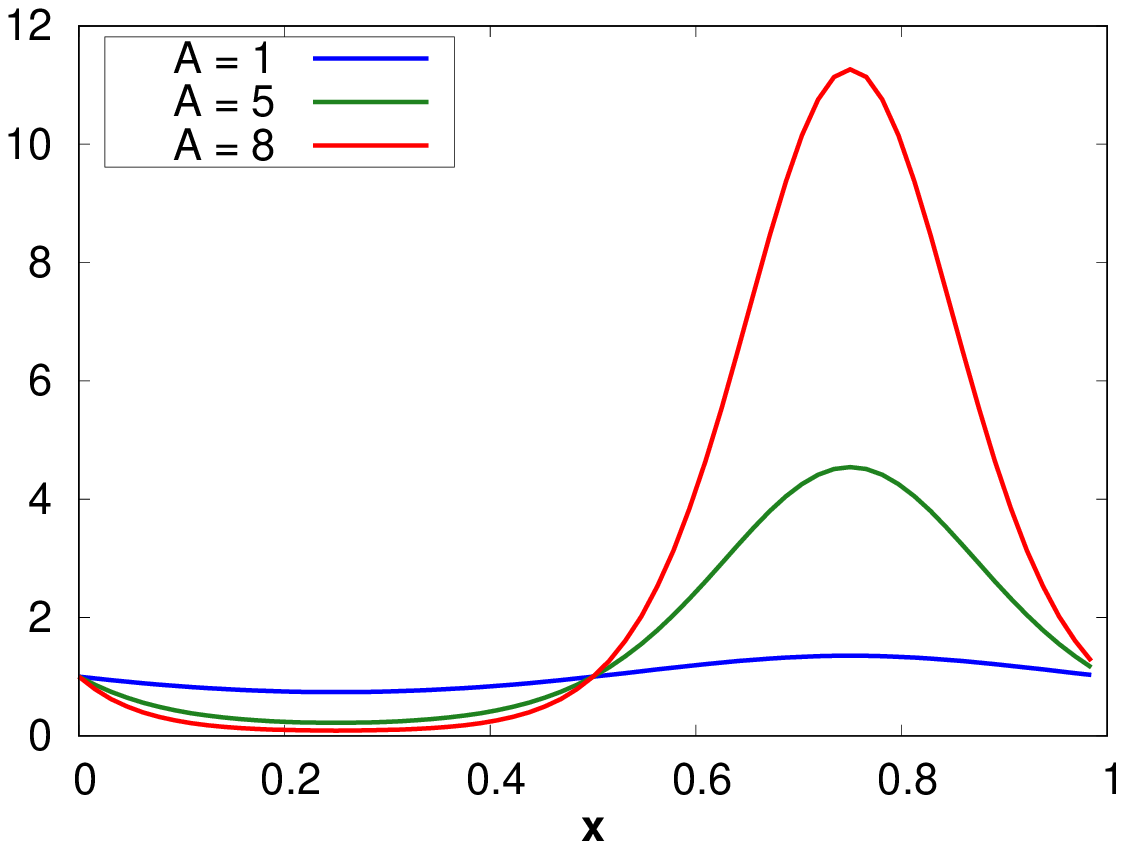}
\caption[A section of $\gamma_{xx}$ for the standing waves test]{\footnotesize{}A section of $\gamma_{xx}$ for the standing waves test. Left: for a small amplitude $A=0.1$, the standing wave is essentially sinusoidal. Right: as the amplitude increases ($A=1,\,5,\,8$) the asymmetry of wave becomes more pronounced. }
\label{a01a10}
\end{figure}

Here we propose a new initial data that satisfies Eq.s~(\ref{constraints}).
Suppose an initial Minkowski flat space, which means zero curvature and where Ricci tensor vanishes ($R=0$).
The constraint equations reduce to 
\begin{subequations} \label{kkk}
\begin{equation} \label{eq:0rezima}
K_{xx}K_{yy}-(K_{xy})^2=0,\\[14pt]
\end{equation} 
\begin{equation}\label{eq:rezima}
D_y K_{xy} - D_x K_{yy} =0,\\[14pt]
\end{equation} 
\begin{equation} \label{eq:primaaa}
D_x K_{xy} - D_y K_{xx} =0.
\end{equation} 
\end{subequations}
The above represent the Hamiltonian and the two momentum constraints, respectively. Note that the Hamiltonian constraint (\ref{eq:0rezima}) requires that the determinant of $K_{ij}$ vanishes. The initial perturbation can be chosen  only for the extrinsic curvature $K_{ij}$, with an unperturbed metric tensor $\gamma_{ij}$. With this choice, one has:
\begin{equation}
\label{stan}
\gamma_{ij}= \eta_{ij}
~,~~~~~	K_{ij}=
\begin{pmatrix}
A sin(2 \pi x) & 0  \\
0 & 0  \\
\end{pmatrix},
\end{equation}
which satisfies the set of constraints in Eq.~(\ref{kkk}).
But unlike the previous gauge waves, this condition does not lead to a wave propagating in space but to standing waves of the metric. 
In Fig.~\ref{a01a10} we report the time evolution of the metric, for several tests.

\begin{figure} [t]
\centering
\includegraphics[width=0.49\textwidth]{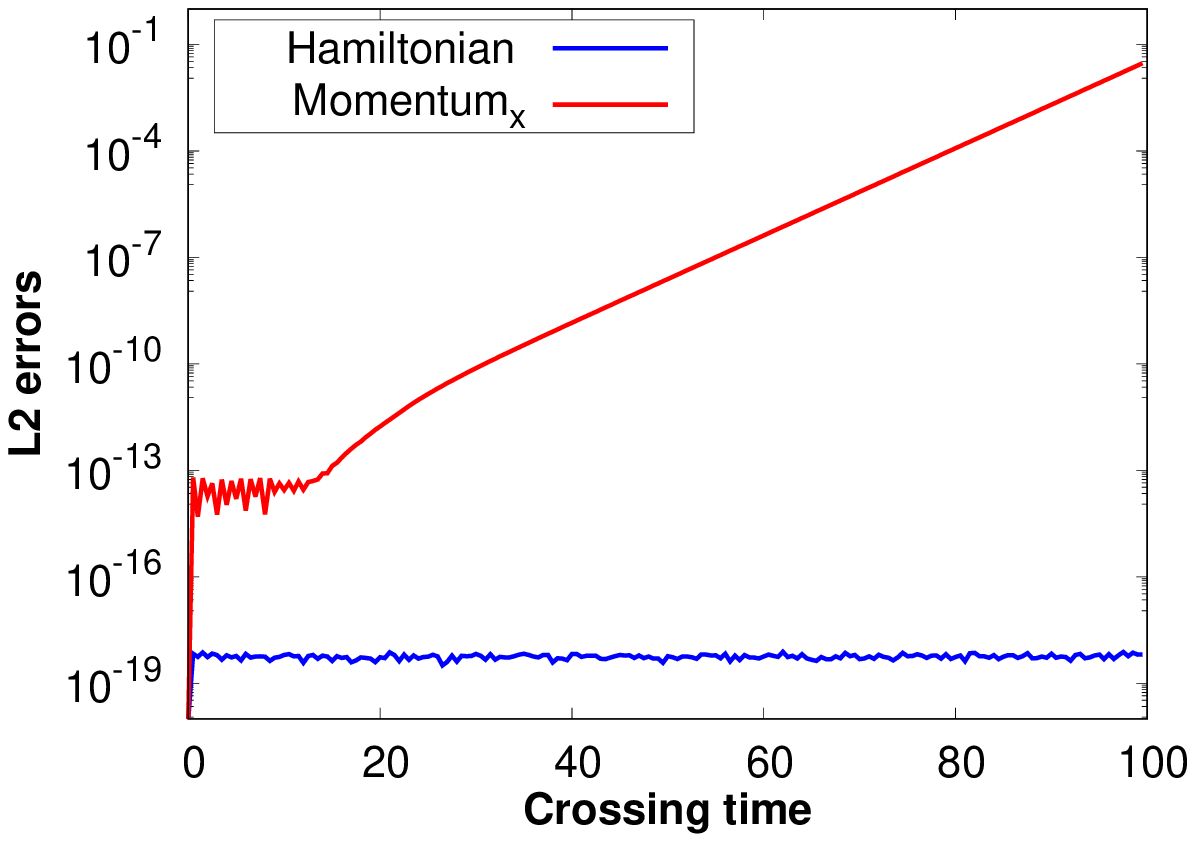}
\includegraphics[width=0.49\textwidth]{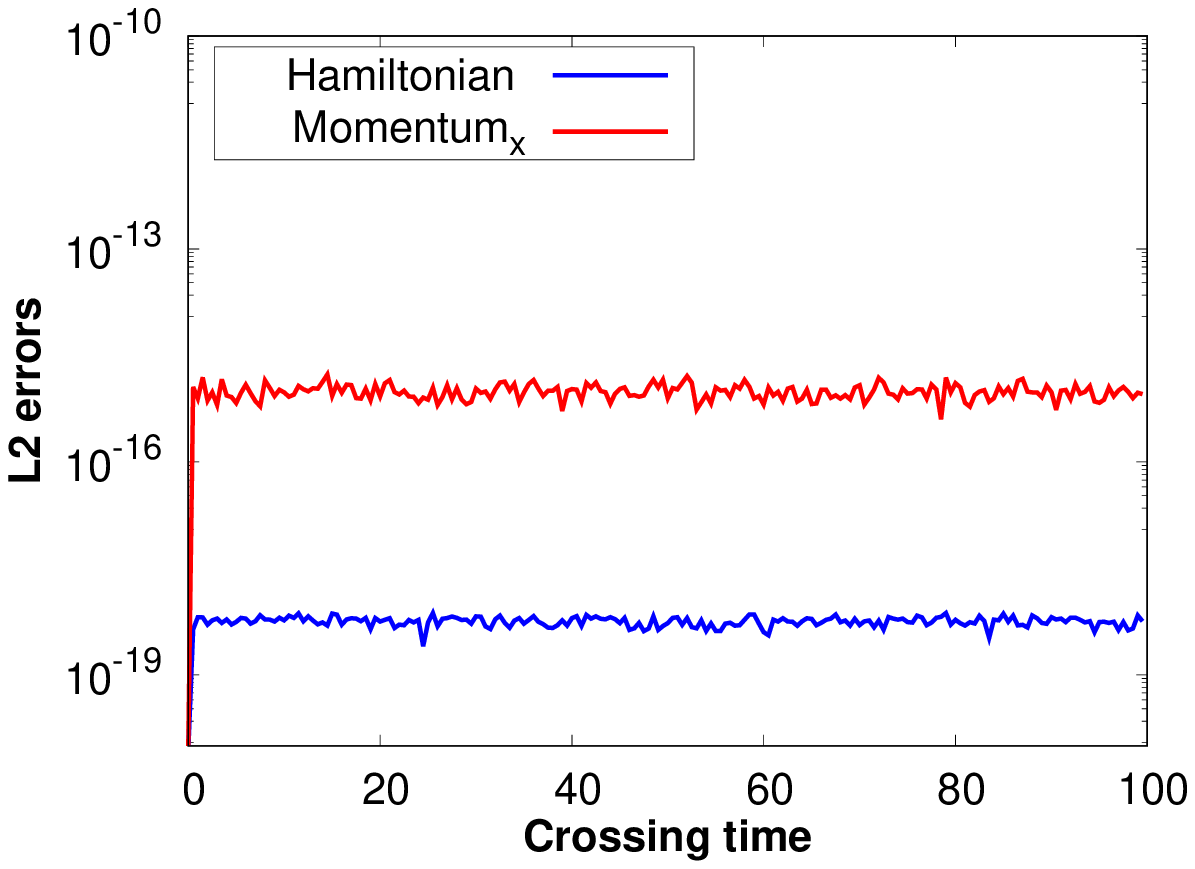}
\caption[]{\footnotesize{}Temporal evolution of the $L_2$ norm of Hamiltonian and $x-$momentum constraint for standing waves test. Left: without filter anti-aliasing: the Hamiltonian constraint remains constant, the $x-$momentum constraint grows up while the $y-$momentum is zero. Right: with $k^* = N/3$ smoothed filter: the Hamiltonian constraint remains essentially unchanged while the $x-$momentum constraint is constant. The evolution is carried out for t = 100 crossing times. These runs are reported in Table~\ref{table} as RUN$_{14}$ and RUN$_{15}$. }
\label{hi100}
\end{figure}

For the first test, a small amplitude $A=10^{-1}$ and no anti-aliasing filter has been used, while the run is carried out until 100 crossing times.
The others parameters used are: a spatial domain $x, y \in [ 0, 1 ]$, a spatial grid $N_x \times N_y = 64^2$, $dx=dy=2^{-6}$, a $dt = 10^{-3}$ and the harmonic slicing ($f(\alpha)=1$).

We have carried out a second test with same parameters as before, except a $k^* = N/3$ smoothed filter has been used.  It is evident that the filter improves the stability. These runs are reported in Table~\ref{table} as RUN$_{14}$ and RUN$_{15}$. Fig.~\ref{hi100} shows that the Hamiltonian constraint remain constant in both cases, but the $x-$momentum constraint (for the symmetry of the problem the $y-$momentum is zero) grows without filter and remains constant with the $k^* = N/3$ filter.

\begin{figure}[t]
\centering
\includegraphics[width=0.65\textwidth]{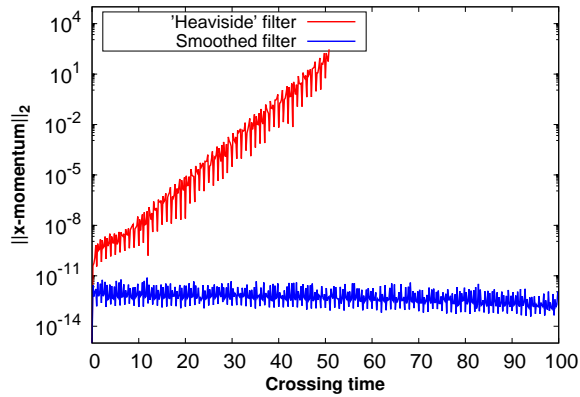}
\caption[]{\footnotesize{}$L_2$ norm evolution of the $x$-constraint with high amplitude $A=8$ using the two different filters described by Eq.s~(\ref{filteralias2}) and (\ref{filteralias1}). The smoothed filter leads to a better stabilized simulation. The evolution is carried out for $t=100$ crossing times. These runs are reported in Table~\ref{table} as RUN$_{16}$ and RUN$_{17}$.}
\label{fdfds}
\end{figure}

In order to see nonlinear effects, a second high amplitude test has been performed, with $A=8$, a spatial grid with $n= 128$ points and a $k^* = N/3$ filter. All the other parameters have been chose as in the previous test.
The higher amplitude of $A$ now induces asymmetry in the standing wave. The metric tensor starts to bounce, and is asymmetric since it cannot become negative on one side. Fig.~\ref{a01a10} (left) shows a section of $\gamma_{xx}$ for a small amplitude $A=0.1$, while in Fig.~\ref{a01a10} (right) amplitude of $A=1,\,5,\,8$ are performed.

Again, this test was performed twice: a first test using the standard truncated filter described by eq.~(\ref{filteralias2}) and a second one in order to emphasize the improvement of the smoothed filter described by Eq.~(\ref{filteralias1}). 
The time evolution of $L_2$ norm of $x-$momentum constraint, computed using the definition in Eq.~(\ref{norml2}), is shown in Fig~\ref{fdfds} for both cases, and the tests are carried out for $100$ crossing times. It is clearly that the smoothed filter works better than the 'Heaviside' filter. These runs are reported in Table~\ref{table} as RUN$_{16}$ and RUN$_{17}$.

As a general summary, finally, in the table \ref{table} we report all the tests performed in this paper. We reported all the main parameters used, such
the type of the initial condition, the amplitude $A$ of the perturbation, the number of point of the spatial grid in $x$ and in $y$, the $k^*$ of anti-aliasing filter, the type of filter and the total time of the run.

\clearpage

\begin{table}
\begin{threeparttable}
\caption{Summary of all the test performed.} \label{table}
\begin{tabular}{lllllll}
\toprule \toprule
\hspace{-3pt}RUN &  \hspace{12pt}IC type  & \hspace{2pt}$A$ & \hspace{-7pt}$ N_x \times N_y$  & \hspace{-1pt} $k^*$  &  \hspace{2pt}filter & $t_{\text{run}}$    \\
\midrule \vspace{3pt}
&     \multicolumn{5}{c}{Standard testbeds}     \\
\vspace{3pt}
~1    & Robust stability test    &$|10^{-10}|$        & $64^2$   &  $\infty$    &  No filter    &  $\sim$ 3   \\   \vspace{3pt}
~2    & Robust stability test    & $|10^{-10}|$         & $64^2$   &  $N/2$       & Heaviside  & ~10  \\ \vspace{3pt}
~3    & Robust stability test    & $|10^{-10}|$         & $64^2$   &  $N/3$       & Heaviside    &  ~10  \\ \vspace{3pt}
~4    & Robust stability test    & $|10^{-10}|$         & $64^2$   &  $N/5 $      &  Heaviside    &  ~10  \\ \vspace{3pt}
~5    & Robust stability test    & $|10^{-10}|$         & $64^2$   & $N/5 $       & Heaviside   &  ~500 \\ \vspace{3pt}
~6    & Robust stability test    & $|10^{-10}|$         & $64^2$   & $N/5 $       & Smoothed   &  ~500 \\ \vspace{3pt}
~7    & Gauge wave test          &$10^{-2}$  & $64^2$   &  $N/3$       &  Heaviside     &  ~$\sim$ 8   \\ \vspace{3pt}
~8    & Gauge wave test          &$10^{-2}$  &$64^2$    &  $N/3$       &  Heaviside   &  ~500 ($dt=5 \cdot 10^{-3}$) \\ \vspace{3pt}
~9    & Gauge wave test          &$10^{-2}$  &$64^2$    &  $N/3$       &  Heaviside   &  ~500 ($dt=10^{-3}$)\\ \vspace{3pt}
~10   & Gauge wave test           &$0.9$    & $128^2$   &  $\infty$   &   No filter      &  ~10  \\ \vspace{3pt}
~11   & Gauge wave test           &$0.9$    & $128^2$   &  $N/3$      &  Smoothed    &  ~10  \\ \vspace{3pt}
~12   & Gauge wave test          &$0.96$   & $128^2$   &  $N/2.5$    & Heaviside    &  ~10  \\ \vspace{3pt}
~13   & Gauge wave test          &$0.96$  & $128^2$   &  $N/2.5$      &  Smoothed  &  ~10  \\ \midrule \vspace{3pt} &    \multicolumn{5}{c}{Our new tests}   \\ \vspace{3pt}
~14   & Standing wave test &$10^{-1}$    & $64^2$   &  $\infty$    & No filter        &  ~100 \\ \vspace{3pt}
~15   & Standing wave test &$10^{-1}$    & $64^2$ &  $N/3$    &   Smoothed  &  ~100 \\ \vspace{3pt}
~16   & Standing wave test &$8$          & $64^2$   &  $N/3$    &  Heaviside   &  ~100 \\
 ~17   & Standing wave test &$8$         & $64^2$   &  $N/3$    & Smoothed    &  ~100  \\
\hline \hline
\end{tabular}
\begin{tablenotes}
\small
\item In this table all the test we performed are summarized. From the left to the right are reported: the number of the run, 
the type of initial condition, the amplitude of the perturbation, the dimension of the spatial grid, the $k^*$ of the anti-aliasing filter, the type of filter and the total time of the run.
\end{tablenotes}
\end{threeparttable}
\end{table}

\section{Conclusions}

In this paper we have studied the evolution of the gravitational fields by solving numerically the Einstein equations.
From the general $3+1$ case, where equations have been decoupled and made more suitable for numerical treatment, we obtained a reduced set of equations in $2+1$ D.
With this formulation it is not possible to simulate a fully gravitational radiation from the merging of compact object such as black holes or neutron stars, but it is however possible simulate standard numerical one-dimensional testbeds.

Our algorithm makes use of a pseudospectral technique for the evaluation of the spatial derivatives. Two types of anti-aliasing filter has been included in order to avoid aliasing instability due to the intrinsic nonlinear nature of the equations. We presented a new technique in order to improve the stability of the simulations named the Running Stability Check (RSC), which monitors the strength of the time derivatives and allows to build also a time-dependent Runge-Kutta step. The code has been written in Fortran and developed entirely by us, from scratch, and successfully tested via classical numerical testbeds of numerical relativity. Via these typical tests, we have proven its robustness and stability.


\begin{thebibliography}{}
%
%


\bibitem{bernuzz4} S. Bernuzzi {\em et al}, {\em Modeling the Dynamics of Tidally Interacting Binary Neutron Stars up to the Merger}, PRL $\bf{114}$, 16 (2014)


\bibitem{bernuz5} S. Bernuzzi {\em et al}, {\em How Loud Are Neutron Star Mergers?}, PRD $\bf{94}$, 024023 (2015)


\bibitem{bernuz3} S. Bernuzzi, A. Nagar, {\em Binary black hole merger in the extreme-mass-ratio limit: A multipolar analysis}, PRD $\bf{81}$, 084056 (2010)


\bibitem{bernuz2} M. Thierfelder {\em et al}, {\em Numerical relativity simulations of binary neutron stars}, PRD $\bf{84}$, 044012 (2011)


\bibitem{camp} M. Campanelli {\em et al}, {\em Accurate Evolutions of Orbiting Black-Hole Binaries without Excision}, PRL $\bf{96}$, 111101 (2006)


\bibitem{naka} T. Nakamura {\em et al}, {\em General relativistic collapse to black holes and gravitational waves from black holes}, Progress of Theoretical Physics Supplement, $\bf{90}$, 1-218 (1987)


\bibitem{rezzol} L. Rezzolla {\em et al}, {\em On the final spin from the coalescence of two black hole}, PRD $\bf{78}$, 044002 (2007)
	
\bibitem{prk} C. Cutler {\em et al}, {\em he Last Three Minutes:  Issues in Gravitational Wave Measurements of Coalescing Compact Binaries}, PRL $\bf{70}$ 2984 (1992)


\bibitem{cap} S. Capozziello, M. Funaro, {\em Introduzione alla relativita' generale. Con applicazioni all'astrofisica relativistica e alla cosmologia}, (book) Liguori Editore, (2006)


\bibitem{ciufolini} I. Ciufolini, V. Gorini, U. Moschella, P. Fr\'e, {\em 
Gravitational Waves (High Energy Physics, Cosmology and Gravitation)}, (book) Institute of Physics Publishing; 1st edition (January 29, 2001)


\bibitem{baumg4} H. Beyer, O. Sarbach, {\em Well-posedness of the Baumgarte-Shapiro-Shibata-Nakamura formulation of Einstein?s field equations}, PRD, $\bf{70}$, 104004 (2004)


\bibitem{admyork} J. W. York {\em Kinematics and dynamics of general relativity}, Cambridge University Press, Cambridge, UK (1979)

\bibitem{baumg3} M. Shibata, T. Nakamura, {\em Evolution of three-dimensional gravitational waves: Harmonic slicing case }, PRD, $\bf{52}$, 5428 (1995)


\bibitem{9} R. P. Kerr,  {\em  Gravitational Field of a Spinning Mass as an Example of Algebraically
	Special Metrics}, PRL $\bf{11}$, 237-238 (1963)


\bibitem{13} R. H. Boyer and R. W. Lindquist {\em Maximal analytic extension of the Kerr metric}, J. Math. Phys. $\bf{8}$, 265-281 (1967)


\bibitem{14} S. Teukolsky {\em The Kerr metric}, Classical and Quantum Gravity $\bf{32}$, (2014)


\bibitem{Bernuz} S. Bernuzzi, D. Hilditch, {\em Constraint violation in free evolution schemes: comparing BSSNOK with a conformal decomposition of Z4}, PRD $\bf{81}$ 084003 (2009)



\bibitem{ADM} R. Arnowitt, S. Deser, C. Misner, {\em Republication of: The dynamics of general relativity}, GRG $\bf{40}$ 1997-2027 (2008)


\bibitem{Alc} M. Alcubierre {\em et al}, {\em Towards standard testbeds for numerical relativity}, Classical Quantum Gravity, $\bf{21}$, 589 (2004)


\bibitem{Dum} M. Dumbser {\em et al}, {\em Conformal and covariant Z4 formulation of the Einstein equations:
Strongly hyperbolic first-order reduction and solution with discontinuous Galerkin schemes}, $\bf{97}$, 084053 (2018)


\bibitem{bab}M. C. Babiuc {\em et al}, {\em Classical Quantum Gravity}, $\bf{25}$, 125012 (2008)


\bibitem{bab2}M. C. Babiuc {\em et al}, {\em Conformal and covariant formulation of the Z4 system with constraint-violation damping}, PRD $\bf{85}$, 064040 (2012)


\bibitem{baumg2} G. B. Cook, {\em Initial Data for Numerical Relativity}, Living Rev. Relativity, $\bf{3}$, 5 (2000)


\bibitem{Brown} J. D. Brown {\em et al}, {\em Numerical simulations with a first order BSSN formulation of Einstein?s field equations}, PRD $\bf{85}$, 084004 (2012)


\bibitem{book} T. Baumgarte, S. Shapiro, {\em Numerical Relativity}, (book) Cambridge University Press (2010)


\bibitem{Alcu} M. Alcubierre, {\em Introduction to $3+1$ numerical relativity}, (book) Oxford Science Publications (2008)


\bibitem{bona1} C. Bona and J. Massó, Phys. Rev. D $\bf{40}$, 1022 (1989)


\bibitem{bona2} C. Bona, J. Massó, E. Seidel, and J. Stela, Phys. Rev. Lett. $\bf{75}$, 600 (1995)


\bibitem{bona3} C. Bona and J. Massó, {\em Republication of: The dynamics of general relativity}, Phys. Rev. Lett. $\bf{40}$, 1022 (1989)


\bibitem{BM} C. Bona, J. Masso', E. Seidel, J. Stela, {\em New Formalism for Numerical Relativity} PRL, $\bf{75}$ 600-603 (1995)


\bibitem{gau4} M. Alcubierre, {\em Hyperbolic slicings of spacetime:  singularity avoidance and gauge shocks}, Classical and Quantum Gravity, 10.1088/0264-9381/20/4/304 (2002)


\bibitem{gau} M. Alcubierre {\em et al},  {\em Gauge conditions for long-term numerical black hole evolutions without excision }, PRD $\bf{67}$, 084023 (2003)


\bibitem{gau3} C. Bona, J. Masso {\em et al}, {\em New Formalism for Numerical Relativity}, PRL $\bf{75}$ 600 (1994)


\bibitem{baumg1} T. Baumgarte, S. Shapiro, {\em Numerical integration of Einstein field equations}, PRD, $\bf{59}$, 024007 (1998)


\bibitem{FFFT} D. Dutykh, {\em A brief introduction to pseudo-spectral methods: application to diffusion problems}. Doctoral. Curitiba, Brazil, pp.55 (2016)


\bibitem{init} E. Gourgoulhon, {\em Construction of initial data for 3+1 numerical relativity},    J. Phys. $\bf{91}$ 012001 (2007)


\bibitem{york1} J. W. York {\em et al}, {\em Gravitational degrees of freedom and the initial-value problem}, PRL $\bf{26}$, 1656-1658 (1971)



\bibitem{ggg}M. M. Boyle {\em et al}, {\em Testing the accuracy and stability of spectral methods in numerical relativity}, PRD $\bf{75}$, 024006 (2007)



\bibitem{kstar} J. P. Boyd, {\em Chebyshev and Fourier Spectral Methods} (book)
Springer-Verlag, Berlin (1989)


\bibitem{canuto} C.Canuto {\em et al}, {\em Spectral Methods in Fluid Dynamics} (book)
Springer Nature, I edition (1 january 1988)



\bibitem{FFT} J. W. Cooley and J. W. Tukey, {\em An algorithm for the machine calculation of complex Fourier series}. Mathematics of Computation, $\bf{19}$ (90) 297 297 (1965)


\bibitem{nrecipes} W. H.Press {\em et al}, {\em Numerical Recipes 3rd Edition: The Art of Scientific Computing}, (book) Cambridge University Press, III edition (September 6, 2007)


\bibitem{CFL} R. Courant, K. Friedrichs e H. Lewy, {\em Uber die partiellen Differenzengleichungen der mathematischen Physik}, Mathematische Annalen, $\bf{100}$, pp. 32-74 (1928)























\end{thebibliography}
\end{document}